\documentclass[aps,physrev,twocolumn,superscriptaddress]{revtex4-2}

\usepackage{graphicx}
\usepackage{dcolumn}
\usepackage{bm}
\usepackage{comment}
\usepackage{appendix}
\usepackage{subcaption}
\usepackage[export]{adjustbox}

\DeclareTextCommandDefault{\textregistered}{\textcircled{R}}

\begin{document}

\title{Multi-Objective Optimizations of High Gradient C-band Photoinjector for High Bunch Charge Applications}

\author{M. Kaemingk}
\email[]{kaemingk@lanl.gov}
\affiliation{Los Alamos National Laboratory}

\author{P. M. Anisimov}
\affiliation{Los Alamos National Laboratory}

\author{J. M. Maxson}
\affiliation{Cornell University}

\author{J. B. Rosenzweig}
\affiliation{University of California, Los Angeles}

\author{E. I. Simakov}
\affiliation{Los Alamos National Laboratory}

\author{H. Xu}
\affiliation{Los Alamos National Laboratory}

\date{\today}

\begin{abstract}
The high gradients potentially achievable in distributed-coupling C-band photoinjectors make them attractive for many high brightness applications. Here we discuss optimization results for a 1.6 cell C-band photoinjector with a 240 MV/m peak field at the cathode that delivers a $250 ~\mathrm{pC}$ electron bunch charge. We use a Multi-Objective Genetic Algorithm (MOGA), obtaining a Pareto front of emittance vs. bunch length. We also perform MOGA optimizations including an aperture to retain only a bright beam core. We find this reduces the emittance of the final beam by more than factor of 2 in some cases. For example, we find that at a root mean square bunch length of 1.6 ps, the use of an aperture improves the transverse emittance from 120 nm to 58 nm assuming negligible photocathode intrinsic emittance. The sacrificial charge at the periphery of the electron beam removed by the aperture linearizes the final slice phase space inside of the remaining beam core. The results obtained surpass the experimental state-of-the-art for beamlines with similar bunch charge.
\end{abstract}

\maketitle

\section{Introduction}
Applications of electron accelerators such as Inverse Compton Scattering (ICS) \cite{Chen2013-MeVXRaysICS, Graves2014-CompactXRayICS} and X-ray Free Electron Lasers (XFELs) \cite{Pellegrini2016-XFELs, Emma2010-FirstLasing, Altarelli2011-EuropeanXFELs, Rosenzweig2020-UCXFEL} have a continuing and profound impact on scientific research in nuclear physics \cite{Schreiber2000-FirstMeasurement}, materials sciences and biology \cite{Chapman2006-FemtosecondDiffractive, Chapman2007-FemtosecondTimeDelay}, medical imaging \cite{Carroll2002-TunableMonochromatic, Weeks1997-ComptonBackscatteringRadiotherapy}, and photolithography processes in the semiconductor industry \cite{Hosler2017-FELlithography}. In these applications, the electron beam brightness is a key figure of merit. Increasing the electron beam brightness leads to improvements in photon production efficiency and reduction of the energy spread of the photon beams produced in ICS and XFELs.

The five dimensional electron beam brightness is defined as
\begin{equation}
    B_{5D} = \frac{Q}{\sigma_L\epsilon_{n,4D}}
    \label{eq:Brightness}
\end{equation}
where $Q$ is the charge, $\sigma_L$ is the root mean square (RMS) length, and $\epsilon_{n,4D}$ is the normalized four dimensional RMS emittance of the electron bunch. Thus increasing the brightness can be achieved by increasing the electron bunch charge, decreasing the beam emittance, and decreasing the bunch length.

The initial beam brightness constitutes an upper bound on the final beam brightness \cite{Bazarov2009-MaxBrightness}. Therefore the quality of the electron source is an important parameter in delivering high brightness beams. Due to their ability to generate short 3D shapeable bunches, photoinjectors are the favored sources for electron beams in these applications. Minimizing the emittance of the photoemitted beam is the subject of ongoing research in photocathode development \cite{Karkare2020-Ultracold, Cultrera2022-PhotoemissionCharacterization, Mohanty2023-MultiAlkaliAntimonide, Martinez-Calderon2024-FabricationAndRejuvenation, Parzyck2023-AtomicallySmooth}. See \cite{Musumeci2018-Advances} for a review of photocathode research for bright beams. The maximum current extractable via photoemission increases as a function of the amplitude of the accelerating field at the photocathode \cite{Bazarov2009-MaxBrightness, Filippetto2014-MaximumCurrentDensity}. Thus operating photoinjectors at high gradients also increases the electron beam brightness. Distributed-coupling cryogenically-cooled copper cavities have been shown to be capable of generating very high fields \cite{Cahill2018-HighGradientCryoCu, Rosenzweig2019-NextGenCryoRF, Rosenzweig2018-UltrahighBrightnessCryoRF, Schneider2022-HighGradientCuCBand, Bai2021-c3}, making them an exciting choice for next generation photoinjectors for high brightness electron beams. Operation at C-band (4-6 GHz) frequencies offers a further advantage over S-band as the higher frequency reduces the power usage and the integrated dark current \cite{Rosenzweig2018-UltrahighBrightnessCryoRF}, while the beam aperture is still large enough for high charge ($> 100~\mathrm{pC}$) bunches.

As the beam propagates, different effects can result in brightness degradation. Variations across longitudinal slices in the evolution of the transverse phase space increase the beam emittance. Additionally, nonlinear forces due to both external fields and internal space charge fields also result in emittance growth, degrading beam brightness. Thus careful tuning of the initial electron bunch and beamline parameters to mitigate and compensate brightness degradation is required. Therefore, optimization studies must be conducted to ensure it is possible for a particular accelerator design to deliver a beam that satisfies the requirements for the desired application. Multi-Objective Genetic Algorithms (MOGA) are well suited for carrying out such optimization studies.

In this paper we discuss the results of optimization studies carried out to explore the achievable brightness for a beam with a bunch charge of $250~\mathrm{pC}$ desired by the cathodes and radio-frequency interactions in extremes (CARIE) project at Los Alamos National Laboratory \cite{Simakov2023-CARIE, Simakov2023-IPACproc}. Following the results in \cite{Li2024-Sacrificial}, we perform optimizations using sacrificial charge. In this scheme, the beam undergoes non-laminar focusing during which the space charge forces of a part of the beam linearize the phase space of a subset of the beam. This results in a dense, bright core which can then be selected out through an aperture. The C-band photoinjector used in these simulations operates with a peak cathode field of $240 ~\mathrm{MV/m}$. This is consistent with experimental work which has demonstrated peak surface fields above $300~\mathrm{MV/m}$ in cold Cu C-band structures for breakdown rates in the $10^{-3}\mathrm{/pulse/m}$ scale \cite{Schneider2022-HighGradientCuCBand}. Previous studies have also considered gradients of $240~\mathrm{MV/m}$ at the cathode in cold copper C-band and even S-band photoinjectors \cite{Robles2021-Versatile, Rosenzweig2018-UltrahighBrightnessCryoRF, Schneider2024-IPACproc}.

\section{\label{Beamline Design} Beamline and Photoinjector Design}

\begin{figure}[htbp]
    \centering
    \begin{subfigure}{0.5\textwidth}
        \includegraphics[width = 0.63\textwidth, left]{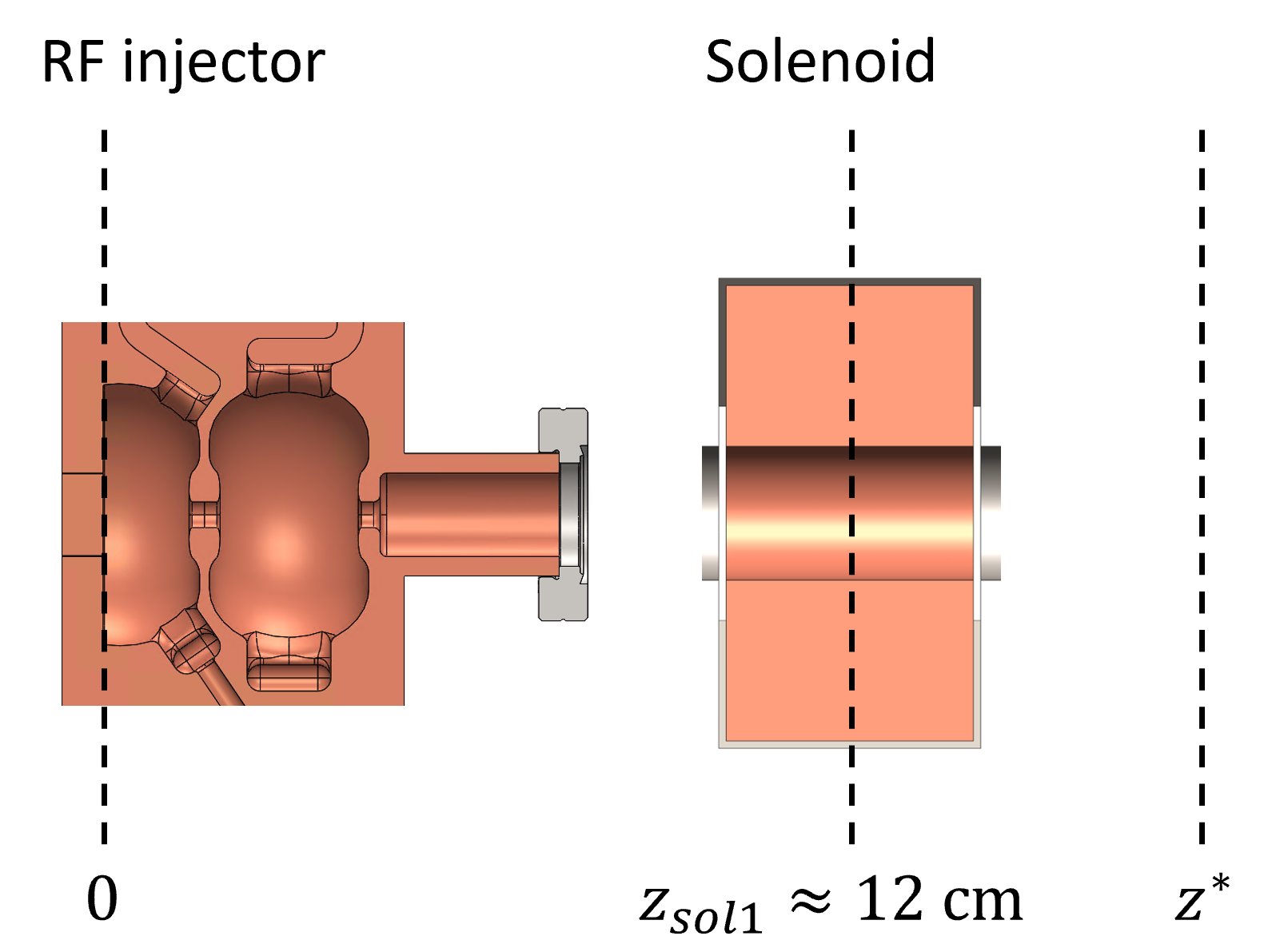}
        \caption{\label{fig:BeamlineNoSacrificial}}
    \end{subfigure}
    \begin{subfigure}{0.5\textwidth}
        \centering
        \includegraphics[width=0.70\textwidth, left]{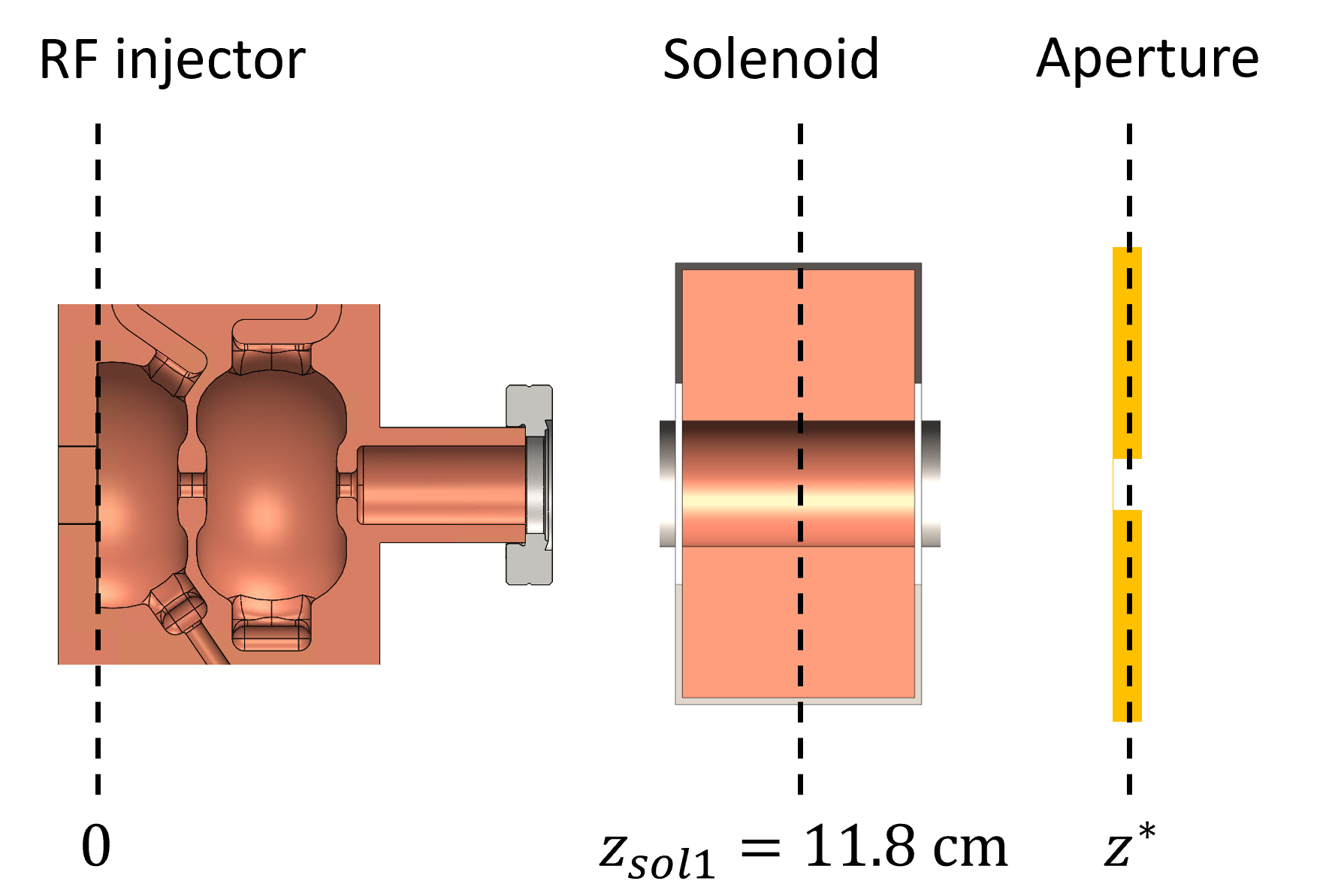}
        \caption{\label{fig:BeamlineSacrificial1Sol}}
    \end{subfigure}
    \begin{subfigure}{0.5\textwidth}
        \centering
        \includegraphics[width=\textwidth, left]{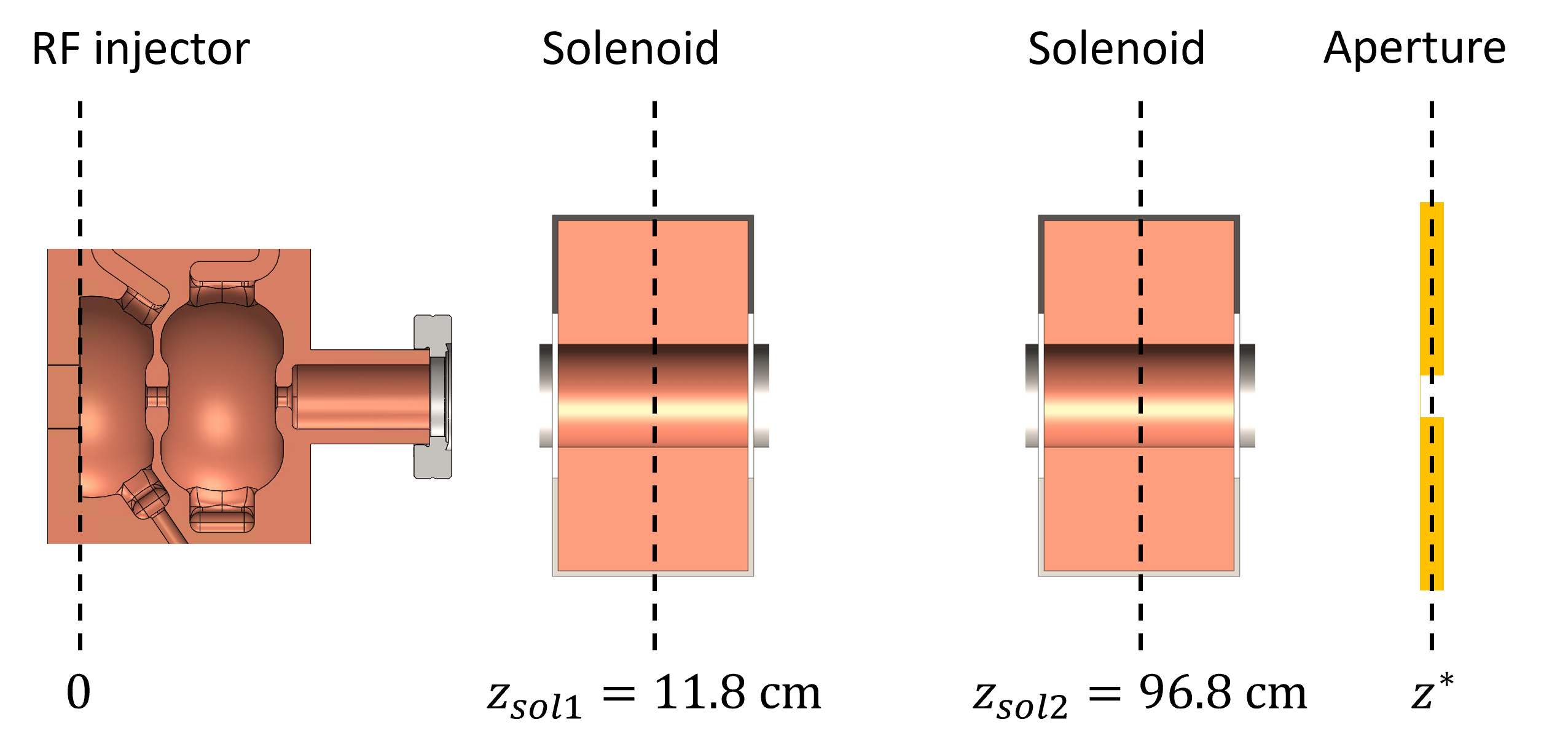}
        \caption{\label{fig:BeamlineSacrificial2Sol}}
    \end{subfigure}
\caption{\label{fig:BeamlineSketch} Sketches of beamline configurations used in the optimizations (a) without sacrificial charge, (b) with sacrificial charge using a single solenoid, and (c) with sacrificial charge using two solenoids. $z^*$ is a variable denoting the location at which the emittance is minimized.} 
\end{figure}

In this section we describe the beamlines and photoinjector designs used in these optimizations. Figure \ref{fig:BeamlineSketch} shows sketches of the different beamline configurations that we simulated.

\begin{figure}
    \centering
    \includegraphics[width=0.4\textwidth]{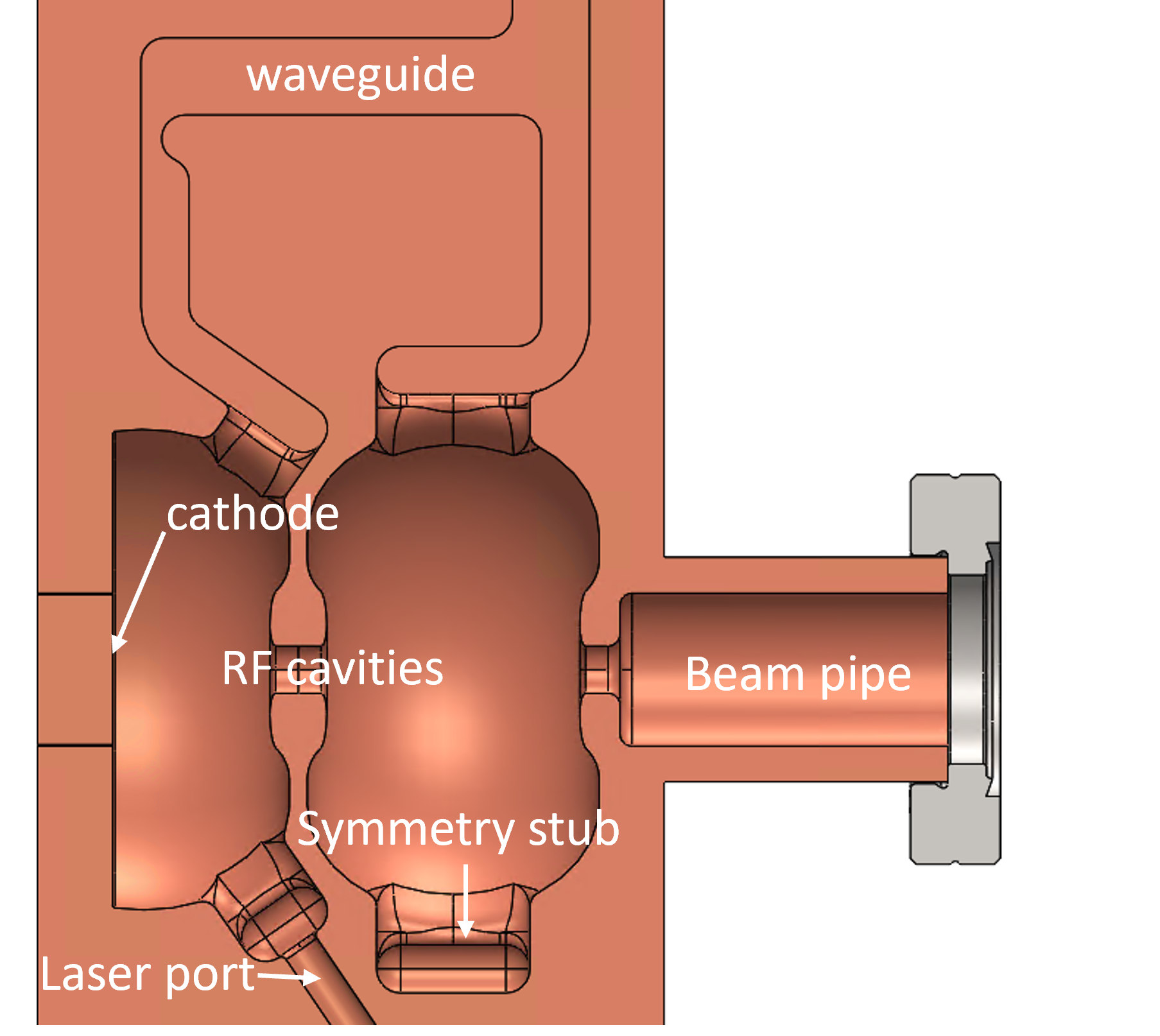}
    \caption{\label{fig:LongGunProfile} Longitudinal photoinjector cross section showing the cavity structures, symmetrizing features, and various ports.}
\end{figure}

The photoinjector is a 1.6 cell 5.712 GHz distributed coupling structure. The original  cavity shape of the 1.6 cell photoinjector was designed at UCLA and is described in \cite{Robles2021-Versatile}. The refined cell profile was designed at SLAC and the coupling waveguide and symmetrization features were designed at LANL\cite{Xu2023-IPACproc}. Figure \ref{fig:LongGunProfile} shows a longitudinal cross section view of the photoinjector. The first cavity is a truncated cell such that it has a flat surface coplanar with the cathode surface.

\begin{figure}[h!]
    \centering
    \begin{subfigure}{0.23\textwidth}
        \centering
        \includegraphics[clip, scale=0.22]{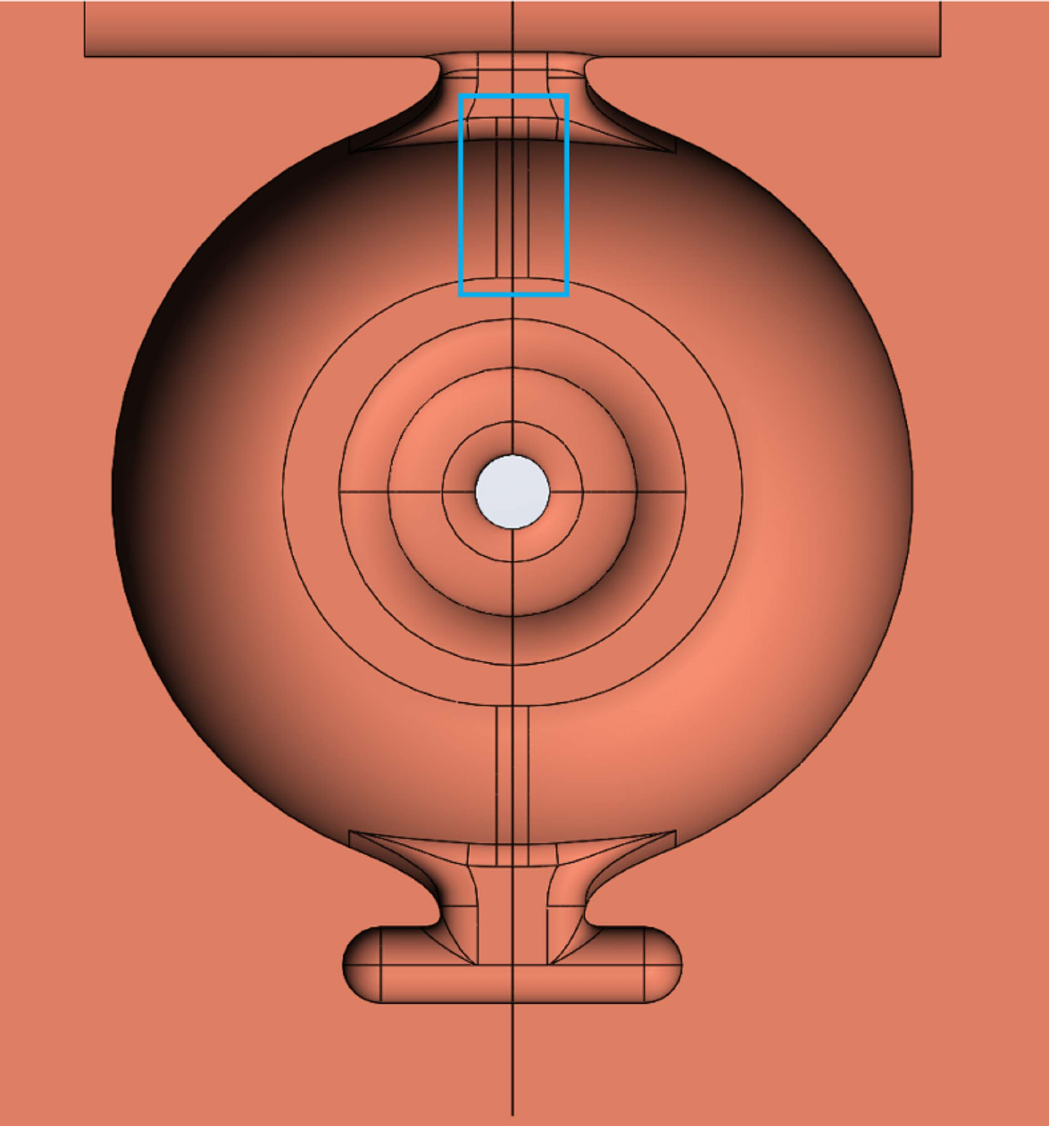}
        \caption{\label{fig:AsymmetricGun}}
    \end{subfigure}
    \begin{subfigure}{0.23\textwidth}
        \centering
        \includegraphics[clip, scale=0.22]{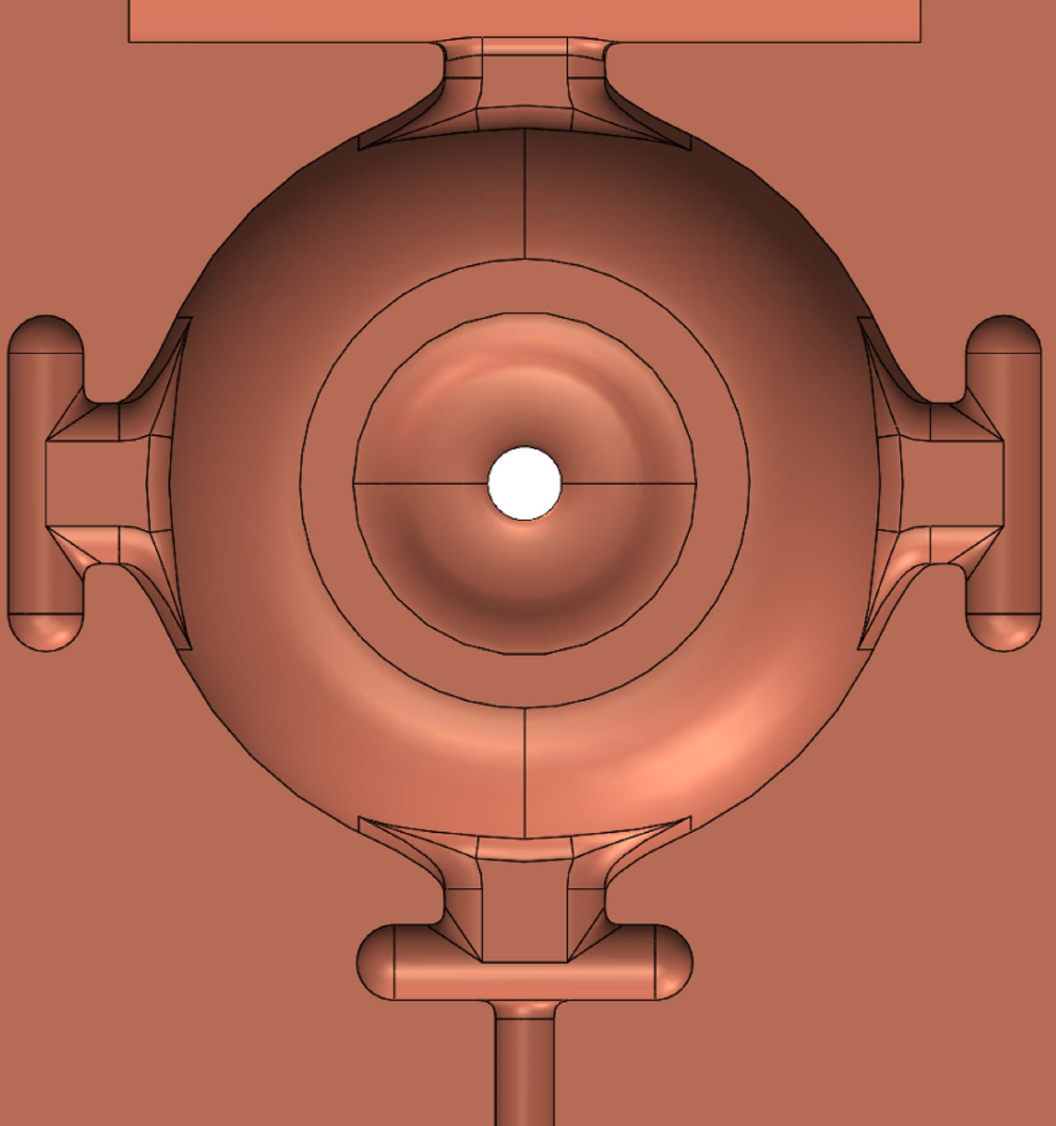}
        \caption{\label{fig:SymmetricGun}}
    \end{subfigure}
\caption{ \label{fig:GunCrossSections} (a) Crossection of initial gun design. The blue rectangle highlights an elongation in the design giving the cross section a ``racetrack" design to mitigate the dipole moment due to the power input coupling ports and the symmetrizing strucutres across from these. (b) Cross section of symmetrized design. The ``racetrack" design has been removed, maintaining a circular cross section, and symmetrizing structures have been added in the direction perpendicular to the power input coupling ports.} 
\end{figure}

The initial cross sectional design of the photoinjector cells is shown in Fig. \ref{fig:AsymmetricGun}. 

\begin{figure}[htbp]
    \centering
    \begin{subfigure}{0.22\textwidth}
        \centering
        \includegraphics[width=\textwidth]{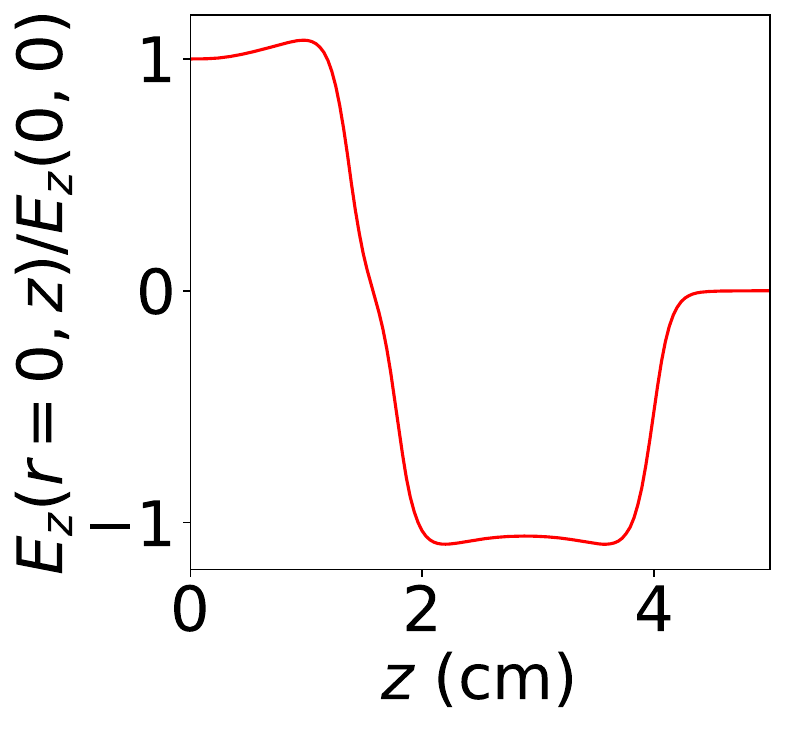}
        \caption{\label{fig:EzProfile}}
    \end{subfigure}
    \begin{subfigure}{0.23\textwidth}
        \centering
        \includegraphics[width=\textwidth]{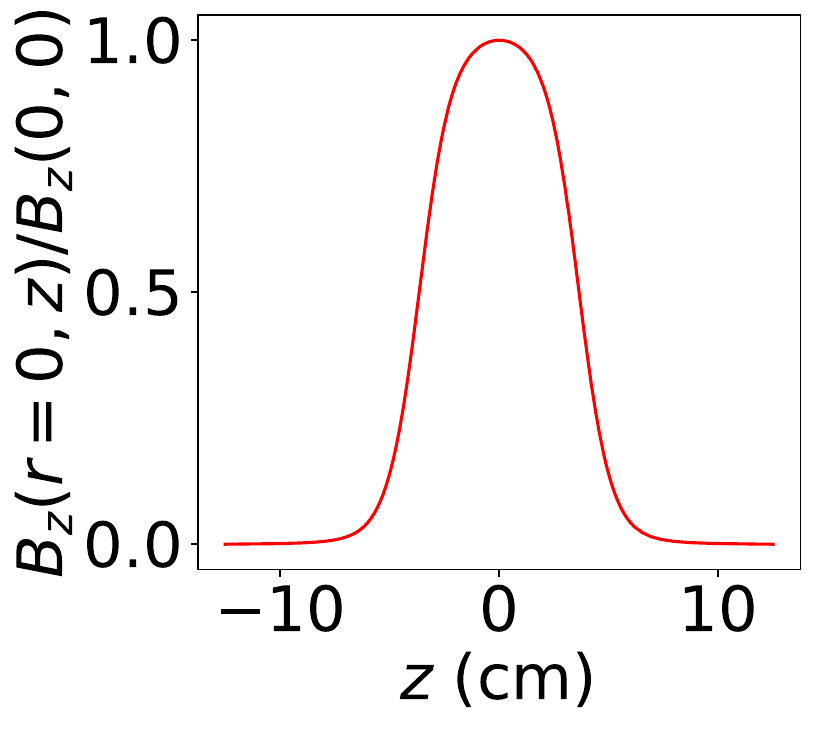}
        \caption{\label{fig:BzProfile}}
    \end{subfigure}
\caption{\label{fig:FieldProfiles} (a) On-axis longitudinal photoinjector electric field profile normalized by the field at the cathode. (b) Longitudinal solenoid field profile normalized by the field at the solenoid center.} 
\end{figure}

In that design there is a single symmetrizing feature directly opposite of the power input coupling port in each cell. Additionally, there is a slight elongation in the perpendicular direction, highlighted by the blue rectangle in Fig. \ref{fig:AsymmetricGun}, giving the cross section a ``race track" design. An important consequence of the optimizations discussed in this paper was the development of a new design for the cross section of the photoinjector cells with increased radial symmetry. This design is shown in Fig. \ref{fig:SymmetricGun} and is discussed in greater detail in \cite{Xu2024-IPACproc, Alexander2024-IPACproc}. The results presented in section \ref{Results} were obtained with the symmetrized design shown in Fig. \ref{fig:SymmetricGun}. In section \ref{Cross Section Comparison} these results are compared with those obtained with the initial design. The photoinjector structure was modeled using SolidWorks\textsuperscript{\textregistered}, and the RF fields were subsequently computed using the CST Studio Suite\textsuperscript{\textregistered} driven-modal solver. Figure \ref{fig:EzProfile} shows the on-axis longitudinal electric field profile of this photoinjector.

The photoinjector accelerates the beam to $\approx6~\mathrm{MeV}$ after which a solenoid located $\approx12~\mathrm{cm}$ from the cathode focuses the beam transversely. Following the work in \cite{Li2024-Sacrificial} which demonstrated the use of sacrificial charge to enhance beam brightness, we performed optimizations with sacrificial charge also including a second solenoid at a distance of $96.8\mathrm{cm}$ away from the cathode. Figure \ref{fig:BzProfile} shows the on-axis longitudinal magnetic field profile of the solenoid used in these simulations. The solenoid was designed at UCLA and is described in \cite{Robles2021-Versatile}.

\begin{figure}[htbp]
    \centering
    \begin{subfigure}{0.5\textwidth}
        \centering
        \includegraphics[width=\textwidth]{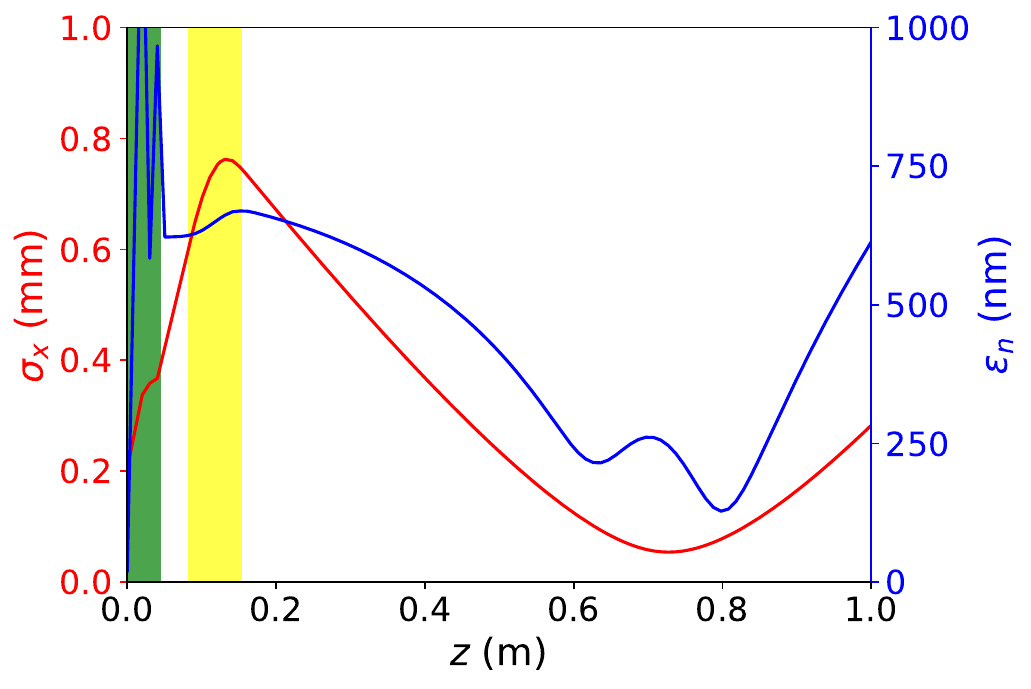}
        \caption{\label{fig:SigmaEmitNS}}
    \end{subfigure}
    \begin{subfigure}{0.5\textwidth}
        \centering
        \includegraphics[width=\textwidth]{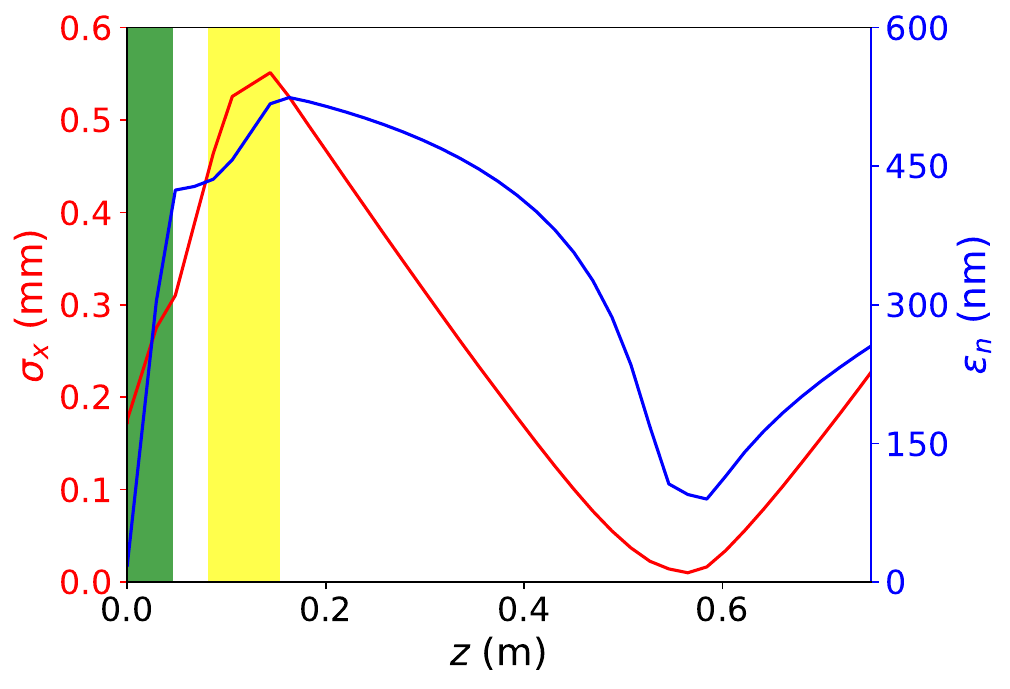}
        \caption{\label{fig:SigmaEmitS1Sol}}
    \end{subfigure}
    \begin{subfigure}{0.5\textwidth}
        \centering
        \includegraphics[width=\textwidth]{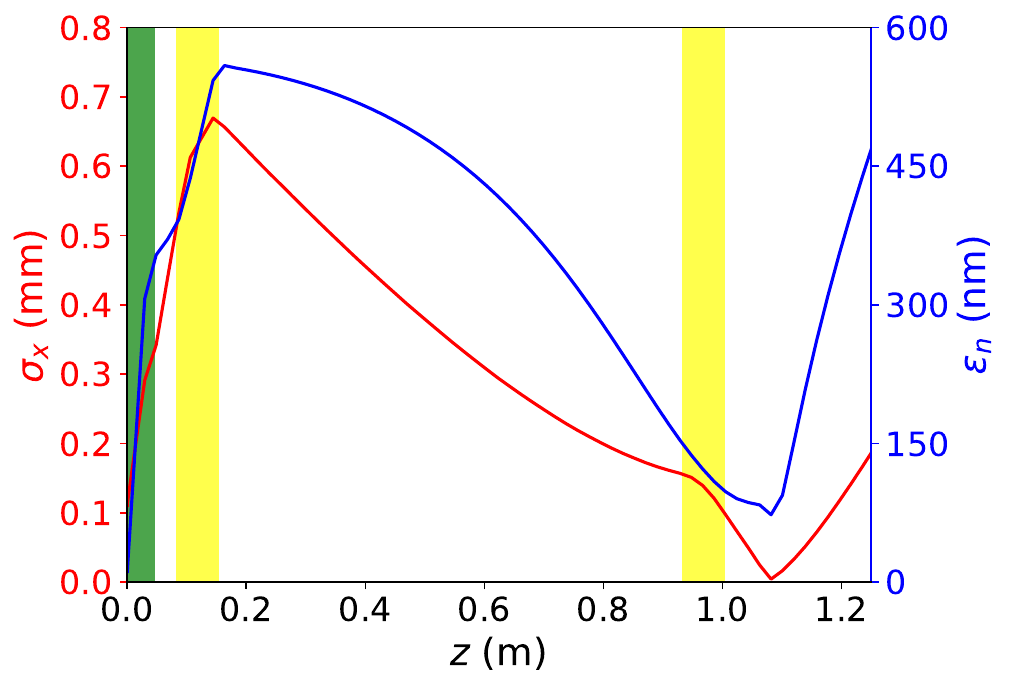}
        \caption{}
        \label{fig:SigmaEmitS2Sol}
        
    \end{subfigure}
\caption{\label{fig:Traces} Example RMS beam size and emittance as a function of distance along the beamline from optimizations (a) without sacrificial charge, (b) with sacrificial charge and one solenoid, and (c) with sacrificial charge and two solenoids. The green shaded region denotes the photoinjector and the yellow shaded regions denote the solenoids.} 

\end{figure}

\section{Optimization Setup}
We performed the simulations using the General Particle Tracer (GPT) code which tracks particles with space charge \cite{ref:gpt_website} and has been well benchmarked against experiments. The number of particles $N_p$ tracked by GPT is a parameter selected by the user which can be increased to improve accuracy and decreased to reduce computational time. Each particle then has charge $Q/N_p$, where $Q$ is the total bunch charge. The RF injector and solenoid fields were converted to GPT data files and loaded into the simulations using built-in GPT functions. 

As seen from equation \ref{eq:Brightness}, for a given bunch charge, the brightness is maximized by minimizing the 4D emittance and the bunch length. Therefore we used a MOGA implemented by Xopt \cite{Xopt:ipac2023-thpl164} to select the simulation settings to minimize the electron bunch 4D emittance and length as discussed below. Each generation in the optimization consisted of $800$ simulations, and the optimizations were concluded after approximately $70$ generations. 

In the subsequent discussion,  the emittance is defined as the square root of the 4D emittance, and is given by
\begin{equation}
    \epsilon_n = \frac{\mathrm{det(<uu^T>)}^{1/4}}{m_e c},
\end{equation}
where $m_e$ is the electron mass, $c$ is the speed of light, $u = (x, p_x, y, p_y)^T$, and the brackets denote an average over the beam distribution.

In each simulation, the emittance is computed at a set of different locations $z_i$ along the beamline to determine the location $z^*$ of the minimum emittance. The optimization objectives are then $\epsilon_{n}(z^*) = \mathrm{min}\{\epsilon_{n}(z_i)\}$ and $\sigma_t(z^*)$. The MOGA optimization yields a Pareto front. This is a curve consisting of points such that moving along the curve to improve in one objective necessarily results in a worsening in another objective. Thus the Pareto front shows the tradeoff among the different objectives. For each point on the Pareto front we compute the brightness according to equation \ref{eq:Brightness}, and determine the point corresponding to the maximum brightness.

The initial beam distribution is given by $\rho(r, t) = Q\rho_\perp(r)\rho_L(t)$. The longitudinal profile of the beam is given by a supergaussian of the form
\begin{equation}
    \rho_L(t) = \frac{1}{2\sqrt{2}\lambda \Gamma(1+
    \frac{1}{2p})}e^{-(t^2/2\lambda^2)^p}.
    \label{eq:InitTDist}
\end{equation}
The supergaussian power $p$ is a parameter that can be varied by the optimizer. In the limit $p\to\infty$, this becomes a uniform distribution. For $p \agt 10$ the distribution already very closely resembles a uniform distribution. The rms length for this distribution is $\sigma_L = \sqrt{\frac{2\Gamma(1+3/2p)}{3\Gamma(1+1/2p)}}\lambda$. The transverse profile is a truncated radial gaussian which has the form
\begin{equation}
    \rho_\perp(r) = \frac{1}{2\pi\sigma^2(1-e^{-n_c^2/2})}\left\{\begin{array}{cc}
         e^{-r^2/2\sigma^2}&  r \leq n_c\sigma\\
         0&  r>n_c \sigma
    \end{array}\right.
    \label{eq:InitRDist}
\end{equation}
The truncation factor $n_c$ is chosen by the optimizer. For $n_c << 1$, this is approximately a uniform radial distribution. The transverse rms size of this distribution is $\sigma_{x,y} = \sigma\sqrt{1-\frac{1}{2} \frac{n_c^2\exp{(-n_c^2/2)}}{1-\exp{(-n_c^2/2)}}}$. These distributions are well approximated by easily achievable laser profiles used for photoemission.

In these simulations we used a half-spherical Gaussian momentum distribution ($p_z > 0$). The RMS momentum spread $\sigma_p$ of this distribution is given by
\begin{equation}
    \sigma_p = \sqrt{m_e K_\perp}
\end{equation}
where $K_\perp$ is the Mean Transverse Energy (MTE). The MTE can be thought of as an effective beam temperature, and in practice is a function of the photoemission wavelength and photocathode work function \cite{Bazarov2008-MTE, Lee2015-UltraLowEmittance, Maxson2015-NaKSbMTE}. In order to study the maximum achievable brightness in the optimized beamlines, a negligible MTE of $3-5~\mathrm{meV}$ was used. This allowed us to isolate the effects of the photoinjector fields, space charge fields, and beam dynamics on brightness degradation from effects due to initial momentum spread.

Table \ref{tab:ParamsAperture} summarizes the optimization variables and ranges used.

\begin{table}[b]
\caption{\label{tab:ParamsAperture}%
Beam and beamline parameters variable by the optimizer. The first column contains the list of parameters and their units when applicable. The second and third columns show the ranges over which the parameters could be varied. Case 1 refers to the case without sacrificial charge. Cases 2 and 3 refer to the cases with sacrificial charge using one and two solenoids respectively. $B_{max}$ is the maximum design field for the solenoids and is approximately $0.65~\mathrm{T}$. The gun phase is measured relative to the maximum momentum gain phase.
}
\begin{ruledtabular}
\begin{tabular}{lcc}
\textrm{Parameters}&
\multicolumn{2}{c}{\textrm{ranges/values}}\\
\colrule
& case 1 & cases 2 \& 3\\
\colrule
Initial $Q$ (pC) & 260 & 250 - 1000\\
Initial $\sigma_{x,y}$ (mm) & 0.01 - 1 & 0.01 - 1\\
$n_c$ & 0.05 - 2 & 0.1 - 3\\
Initial $\sigma_L$ (ps) & 0.5 - 2.5 & 0.5 - 50\\
$p$ & 1 - 100 & 1 - 100\\
Gun phase (deg) & -25 - 25 & -15 - 15\\
$B_{sol1}/B_{max}$ & 0 - 1.3 & 0 - 1.2\\
$z_{sol,1}$ (cm) & 11.5 - 13.5 & -\\ 
$B_{sol2}/B_{max}$ (when used) & - & 0 - 1\\

\end{tabular}
\end{ruledtabular}
\end{table}
\section{\label{Results} Results}
\begin{figure}[htbp]
\centering
\includegraphics[width=0.5\textwidth]{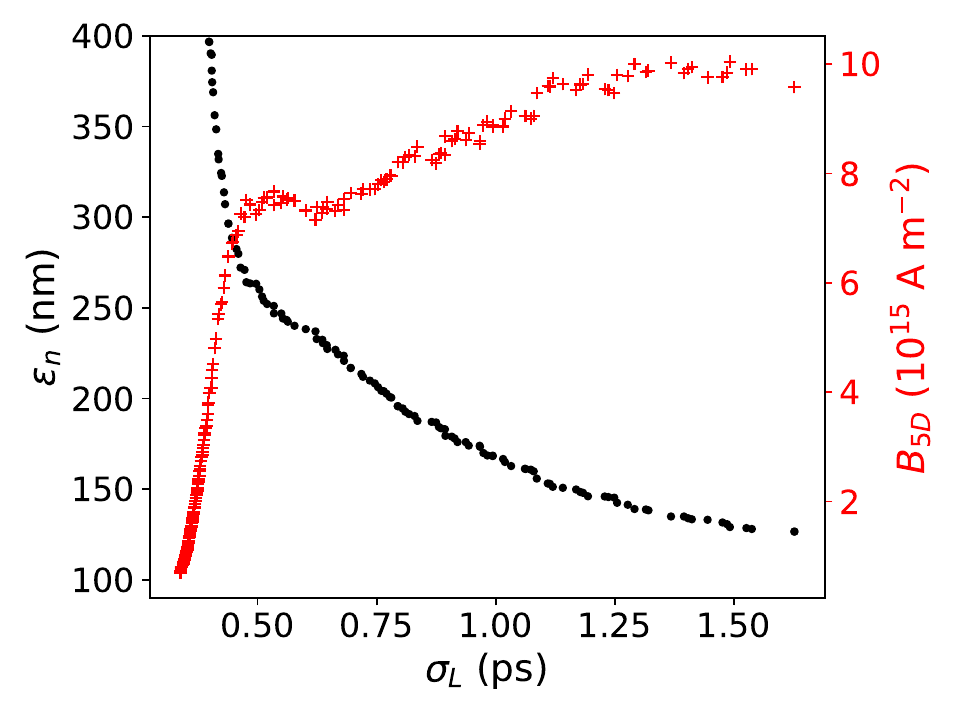}
\caption{\label{fig:NoApertureFrontFinal} Pareto front from MOGA optimizations without sacrificial charge. The brightness is computed from the emittance and bunch length according to equation \ref{eq:Brightness}.}  

\end{figure}

In this section we report the results from the optimizations described in the previous section. Figure \ref{fig:Traces} shows examples of the variation of the emittance and bunch width for the three optimized cases. Figure \ref{fig:NoApertureFrontFinal} shows the Pareto front obtained without using sacrificial charge. We refer to these as case 1. 
For these simulations the number of particles $N_p$ was 30000 and the total bunch charge was $260~\mathrm{pC}$, chosen to obtain a final bunch charge of $250~\mathrm{pC}$. An initial MTE of $5~\mathrm{meV}$ was used. The location of the first solenoid $z_{sol,1}$ was an optimization variable and only one solenoid was used. 

\begin{figure}[htbp]
    \centering
    \begin{subfigure}{0.5\textwidth}
        \centering
        \includegraphics[width=\textwidth]{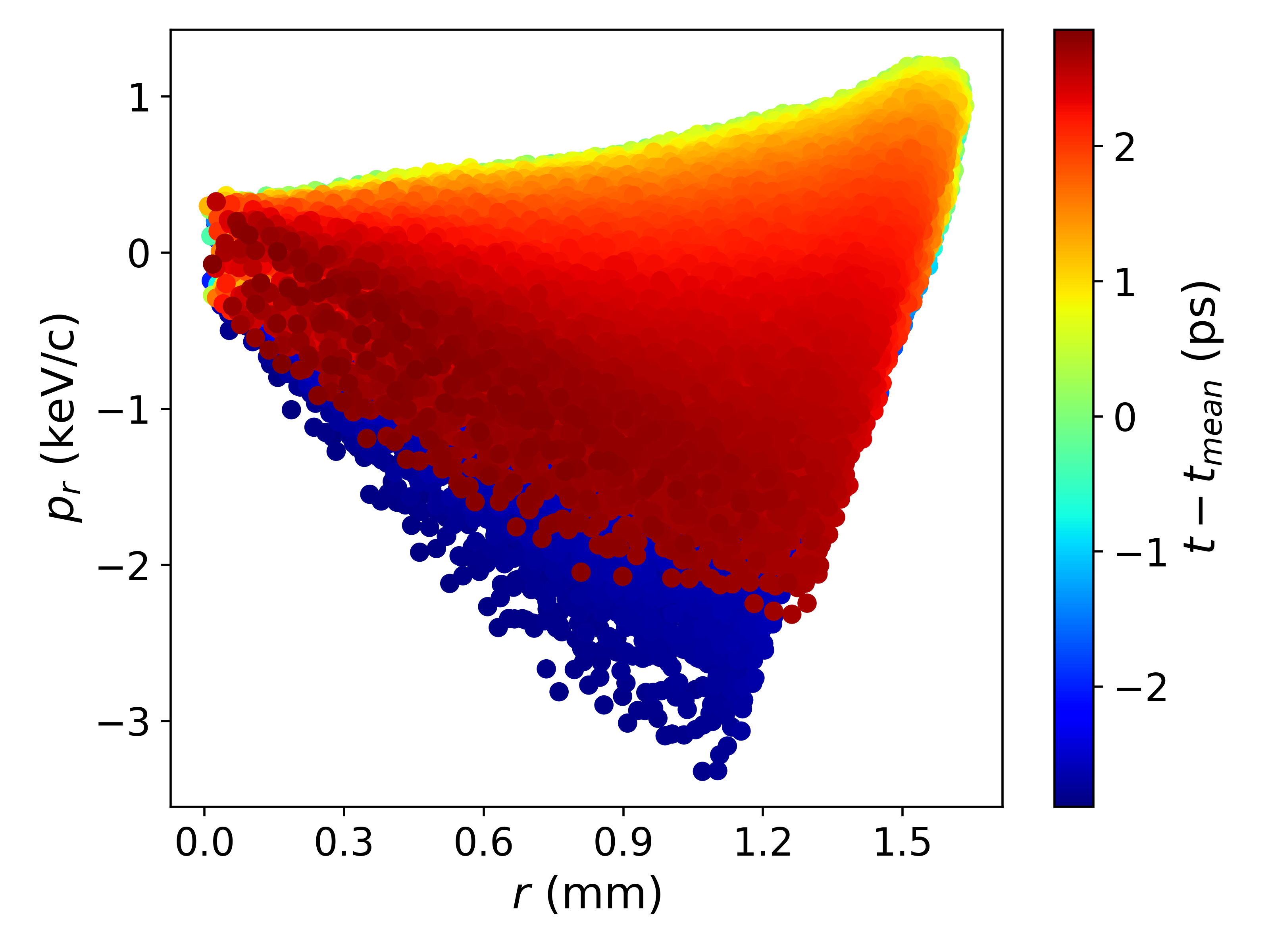}
        \caption{\label{fig:rpr_sol}}
    \end{subfigure}
    \begin{subfigure}{0.5\textwidth}
        \centering
        \includegraphics[width=\textwidth]{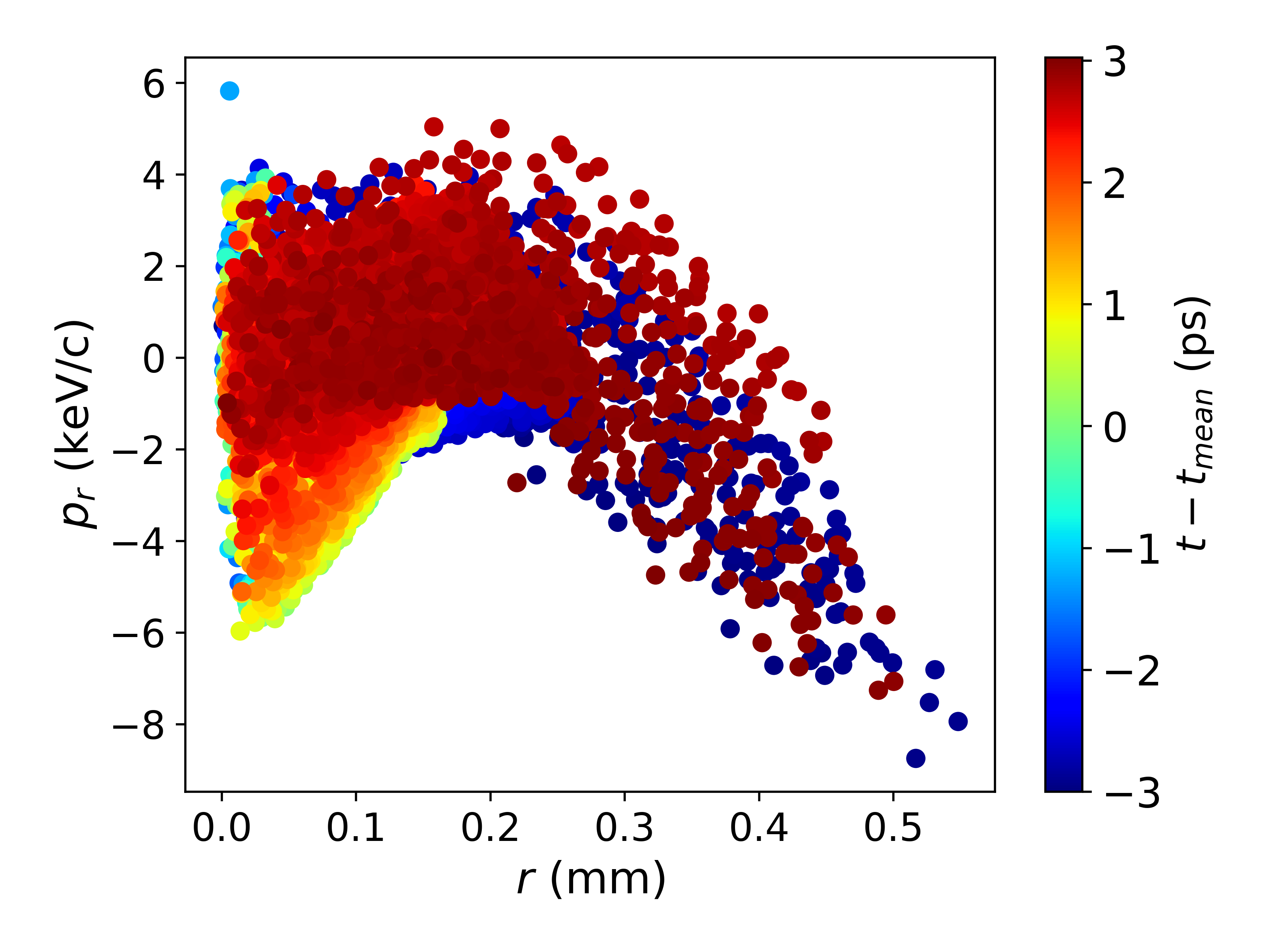}
        \caption{\label{fig:rpr_merit}}
    \end{subfigure}
\caption{\label{fig:PhaseSpaceNoAperture} Phase spaces from example from optimizations performed without sacrificial charge. (a) Phase space at the location of the solenoid. (b) Phase space at the location where the emittance is minimized.}
\end{figure}

From these optimizations we found a minimum emittance of $127~\mathrm{nm}$ at a bunch length of $\approx 1.6~\mathrm{ps}$ and a maximum 5D brightness of $10^{16} ~\mathrm{A/m^2}$ at a bunch length of $\approx 1.5~\mathrm{ps}$. In all cases the optimizer selected a supergaussian power $p > 20$ and $n_c<0.5$, favoring a uniform profile both longitudinally and transversely. Figure \ref{fig:PhaseSpaceNoAperture} shows the $r-p_r$ phase spaces for the minimum emittance case at the location of the solenoid and at the location at which the emittance is minimized. The points are colored by their longitudinal position along the bunch quantified by $t-t_{mean}$, where $t$ is the time at which each point arrives at the reference location and $t_{mean}$ is the average of these times. Particles at the front of the beam arrive earlier, implying a more negative value for $t-t_{mean}$, than those further back.
We see a dependance of the radial momentum on longitudinal position at the solenoid. This dependence is greatly diminished at the location of minimum emittance, in agreement with the well known emittance compensation mechanism due to slice realignment \cite{Serafini1997-EmittanceComp}. However a clear nonlinearity can be observed in the radial phase space which prevents further emittance decrease. To address this issue, we performed optimizations using sacrificial charge to investigate the application of the radial phase space linearization scheme reported in \cite{Li2024-Sacrificial} to our beamline and desired bunch charge.

In these optimizations, the initial bunch charge is a variable parameter to be optimized. Prior to computing the transverse emittance, at each location along the beamline, the particles within some radius are selected. This radius is chosen so that the charge of the selected particles is the desired final $250~\mathrm{pC}$. Physically this represents the beam passing through an aperture. The transverse emittance of the selected particles is computed to determine the location that minimizes this emittance. The minimization objectives are then the minimum emittance and bunch length of the selected particles, denoted the ``survivors". The particles that are not selected are denoted the ``sacrificial" particles.

We performed optimizations using sacrificial charge and a single solenoid to explore the maximum brightness we could obtain in that case. This is referred to as case 2. We additionally performed optimizations with sacrificial charge and a beamline using two solenoids as was done in \cite{Li2024-Sacrificial}. This is referred to as case 3. The first solenoid was located $0.118~\mathrm{m}$ downstream of the cathode. An MTE of $3~\mathrm{meV}$ was used in these optimizations.

\begin{figure}[htbp]
\centering
\includegraphics[width=0.5\textwidth]{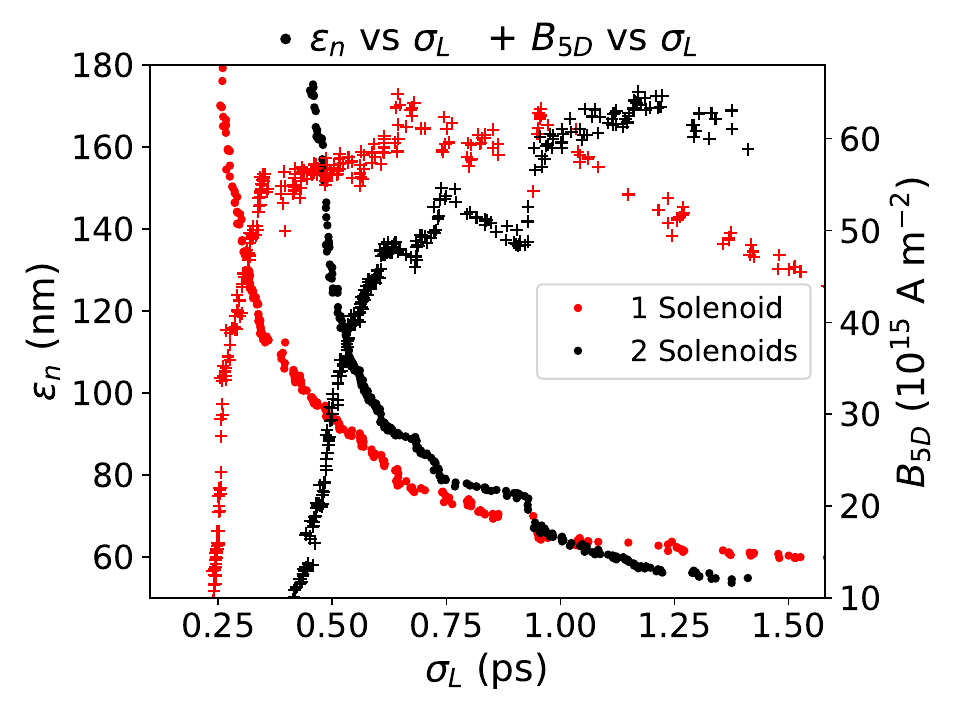}
\caption{\label{fig:1sol2solFronts} Pareto fronts from MOGA optimizations using sacrificial charge. The brightness is computed from the emittance and bunch length according to equation \ref{eq:Brightness}.}
\end{figure}
The Pareto fronts obtained for both cases 2 and 3 are shown in Fig. \ref{fig:1sol2solFronts}.

\begin{figure}[h!]
\centering
\begin{subfigure}{0.5\textwidth}
    \centering
    \includegraphics[scale=0.55]{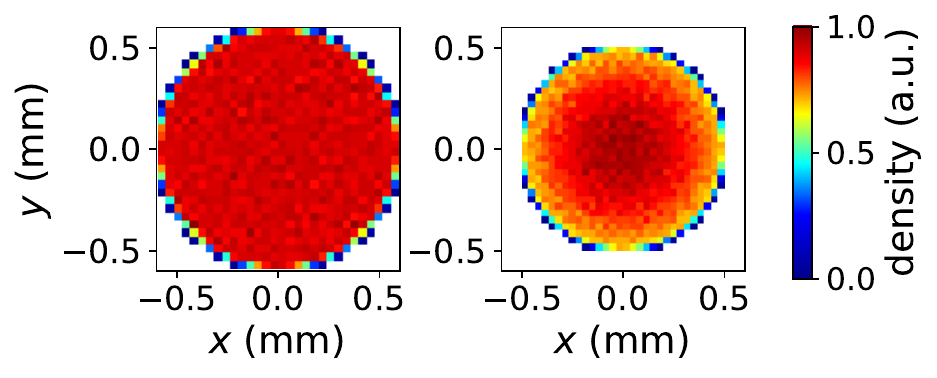}
    \caption{\label{fig:profiles}}
\end{subfigure}
\begin{subfigure}{0.5\textwidth}
    \centering
    \includegraphics[scale = 0.5]{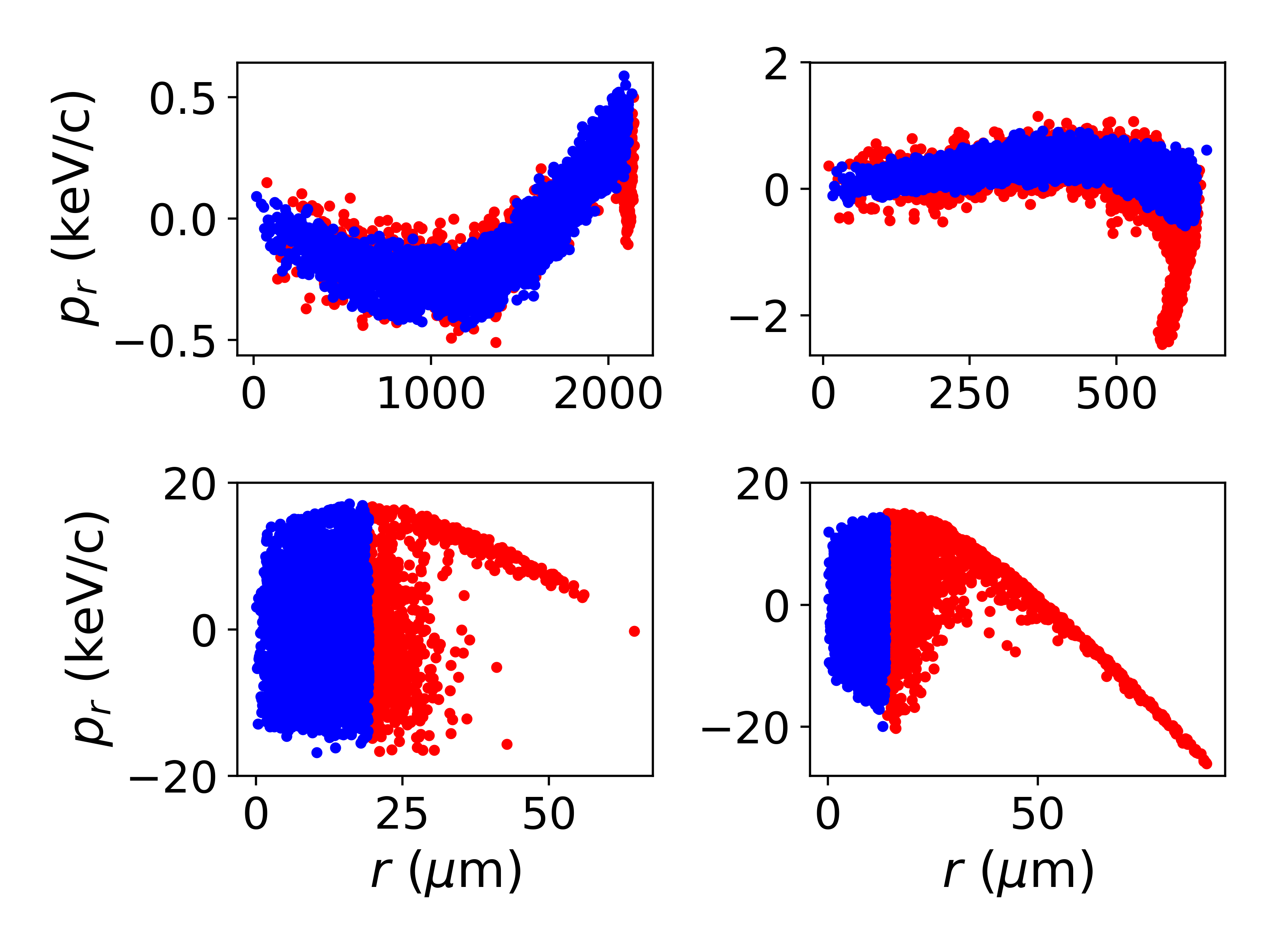}
    \caption{\label{fig:phasespaces}}
\end{subfigure}
\caption{\label{fig:PhaseSpace} Examples of the initial beam distribution and radial slice phase spaces from optimizations using sacrificial charge. (a) Initial transverse beam distribution. (b) Radial phase spaces of the middle slice at the location of the last solenoid (top) and at the location where the transverse emittance of the survivors is minimized (bottom). Blue points correspond to the survivor particles, and red points correspond to the sacrificial particles. In both (a) and (b) the left plots are from optimizations with a single solenoid and the right plots are from optimizations with two solenoids.}
\end{figure}

In these simulations, the number of survivor particles was 20000, and we found the number of survivors to be approximately the same as the number of sacrificial particles. As in the optimizations without sacrificial charge, the optimizer favored a longitudinally uniform profile in all cases. We found that the use of sacrificial charge significantly increases the achievable brightness. In both cases 2 and 3, we obtained a maximum brightness of $\approx6.5 \times 10^{16}~\mathrm{A/m^2}$. In case 2, with a single solenoid, the maximum brightness is achieved at a bunch length of $\approx0.6~\mathrm{ps}$. For higher bunch lengths, the brightness decreases. This is because for greater bunch lengths, the rate at which the bunch length increases along the Pareto front is faster than the rate at which the 4D emittance decreases. In case 3, the maximum brightness is achieved at a longer bunch length of $\approx 1.2 ~\mathrm{ps}$. Although both cases yielded nearly identical maximum brightness, the initial distributions and resulting dynamics show significant differences. Figure \ref{fig:profiles} shows an example of the initial distributions. Figure \ref{fig:phasespaces} shows the slice radial phase spaces observed in each case illustrating the different dynamics. To highlight the different dynamics, we selected simulation settings which resulted in very similar survivor beam brightness at similar survivor bunch lengths of approximately $1~\mathrm{ps}$.

In case 3, we see that the initial transverse beam distribution favored by the optimizer is a radial Gaussian distribution truncated at a radius close to $\sigma$, the standard deviation of the Gaussian distribution. For example in the case shown in Fig. \ref{fig:PhaseSpace}, the truncation radius is $0.83\sigma$. As the beam propagates to the second focusing solenoid, the space charge fields of this distribution cause the slice radial phase spaces to develop a concave downward shape. This results in a radial shell of higher charge density forming around each slice. As discussed in \cite{Li2024-Sacrificial}, as the beam is nonlaminarly focused by the second solenoid, the space charge fields of the shell apply an impulse to the survivors which linearizes the phase space of the survivor particles.

In case 2, the initial distribution has a truncation radius of approximately $0.1\sigma$ producing an approximately radially uniform distribution. The shell formed in this case has much smaller density and transverse velocity spread compared to that of the two solenoid case. However, the beam is also non-laminarly focused after which the radial phase space of the survivor particles is found to be linearized.

\begin{figure}[h!]
    \centering
    \begin{subfigure}{0.5\textwidth}
        \centering
        \includegraphics[trim={0 1.5cm 0 1.5cm},clip, scale=0.5]{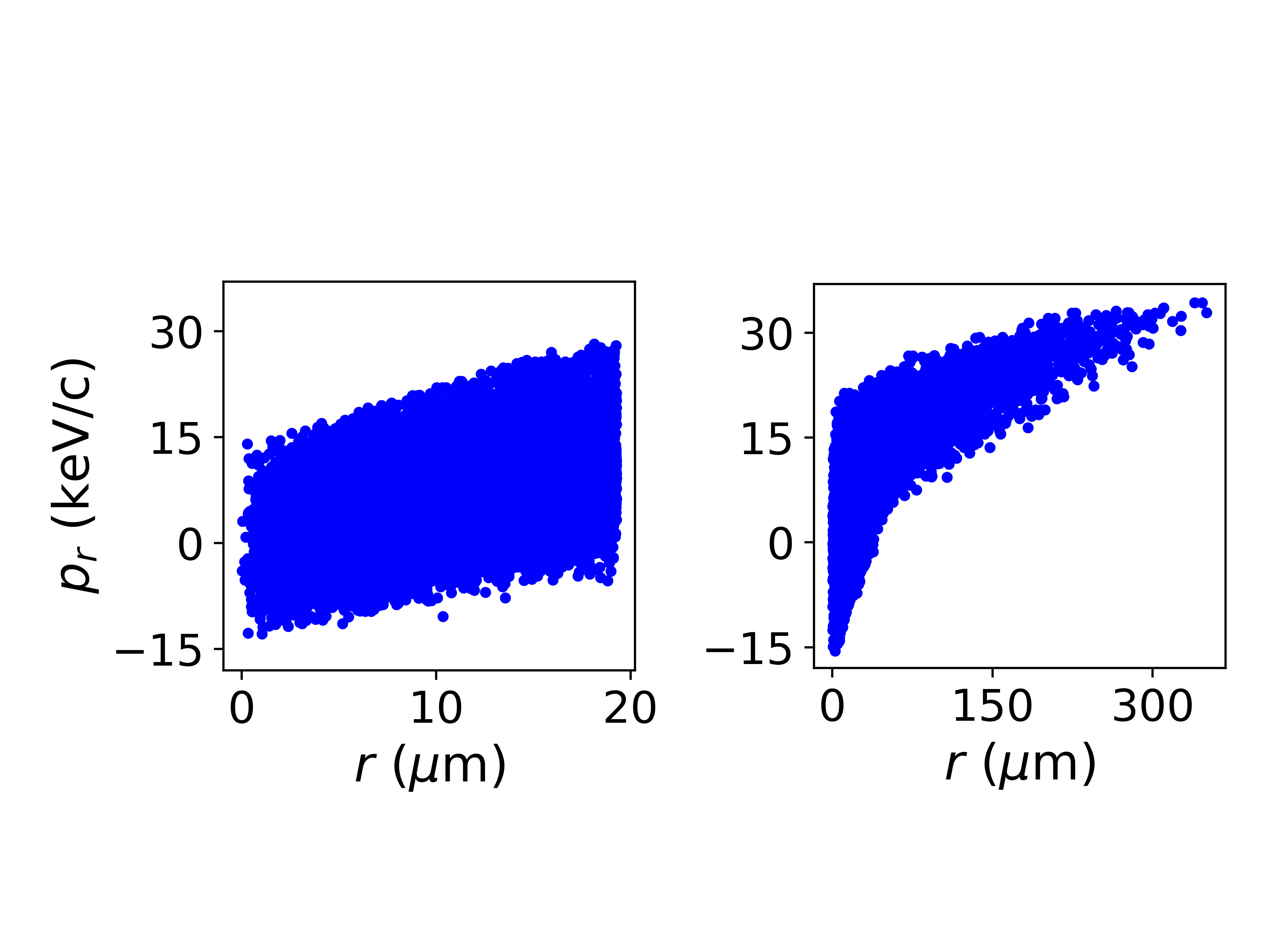}
        \caption{\label{fig:phasespace1sol}}
    \end{subfigure}
    \begin{subfigure}{0.5\textwidth}
        \centering
        \includegraphics[trim={0 1.5cm 0 1.5cm}, clip, scale=0.5]{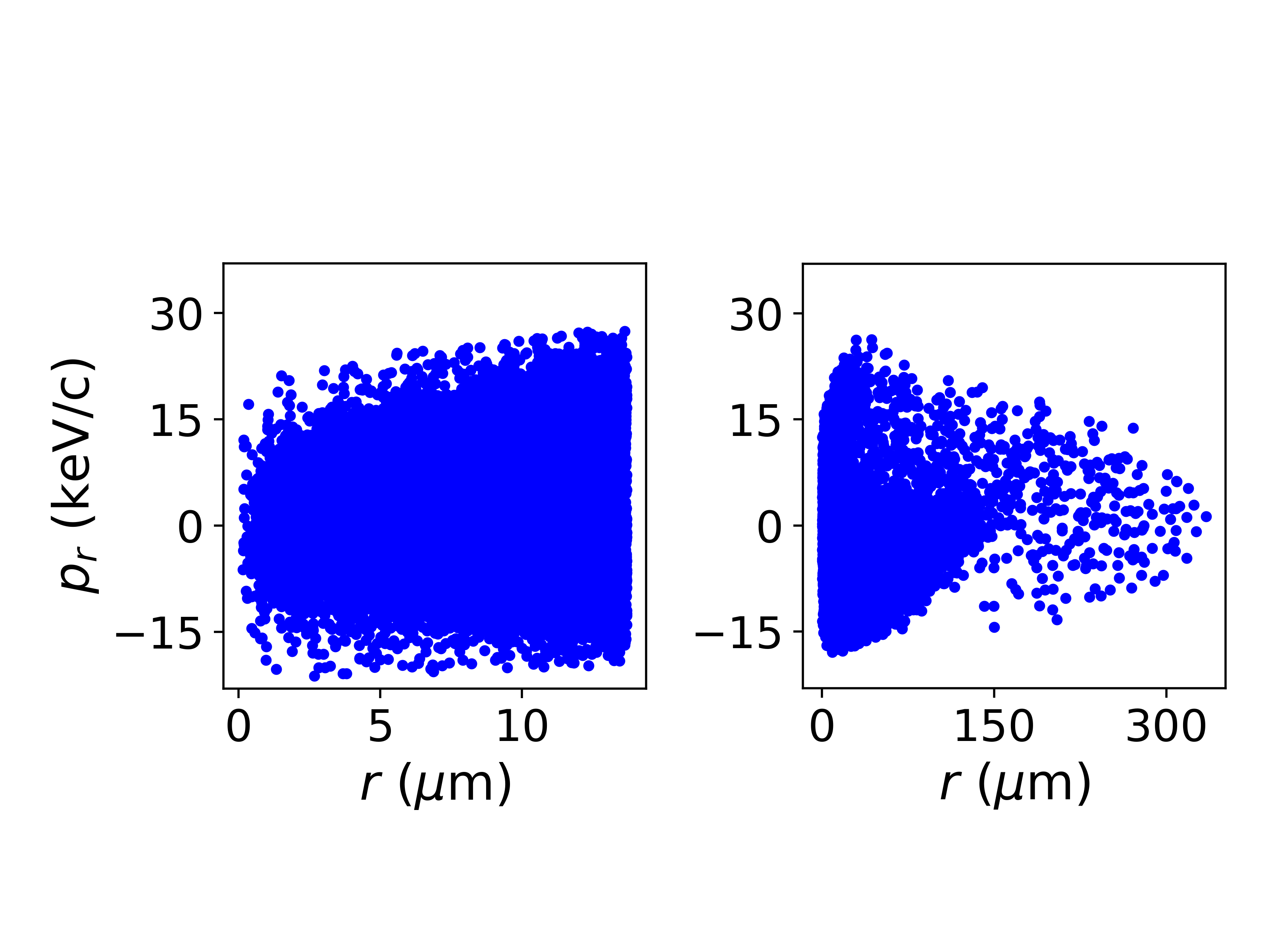}
        \caption{\label{fig:phasespace2sol}}
        
    \end{subfigure}
\caption{\label{fig:1sol2solPhaseSpaces} Example radial phase spaces of the survivors at the location where the survivor emittance is minimized. (a) Example from the one solenoid case. (b) Example from the two solenoid case. In both (a) and (b) the left plot shows the phase space when propagated with sacrificial charge and the right plot shows the phase space when propagated without sacrificial charge.} 
\end{figure}

\begin{table}[b]
\caption{\label{tab:Settings}%
Beam and beamline parameter ranges for simulations yielding emittances within $20\%$ of the minimum emittances. In the case of the supergaussian power $p$, values above $20$ all give temporal profiles whose differences are negligible.
}
\begin{ruledtabular}
\begin{tabular}{lccc}
\textrm{Parameters}&
\multicolumn{3}{c}{\textrm{Ranges}}\\
\colrule
& case 1 & case 2 & case 3\\
\colrule
Initial $Q$ (pC) & 260 & 480 - 573 & 475 - 575\\
Initial $\sigma_{x,y}$ (mm) & 0.21 - 0.27 & 0.25 - 0.35 & 0.15 - 0.25\\
$n_c$ & 0.05 - 0.30 & 0.10 - 0.26 & 0.72 - 0.89\\
Initial $\sigma_L$ (ps) & 1.2 - 1.7 & 1.2 - 2.9 & 1.4 - 2.8 \\
$p$ &  $\geq$ 20 & $\geq$ 20 & $\geq$ 20\\
Gun phase (deg) & -4.4 - -2.1 & -3.4 - 0.6 & -3.2 - 0.4\\
$B_{sol1}/B_{max}$ & 0.98 - 1 & 1.01 - 1.07 & 0.97 - 0.98\\
$z_{sol,1}$ (cm) & 11.5 - 11.9 & - & -\\ 
$B_{sol2}/B_{max}$ & - & - & 0.97 - 0.98\\

\end{tabular}
\end{ruledtabular}
\end{table}

To observe the impact of the sacrificial charge in the linearization process, we propagated the survivor particles from the last solenoid without the sacrificial charge. Figure \ref{fig:1sol2solPhaseSpaces} shows the radial phase spaces of the survivor particles at the location where the survivor emittance is minimized. Shown are the radial phase spaces when the particles are propagated with and without sacrificial charge for both cases 2 and 3. In both instances, we see the phase spaces after propagating without the sacrificial charge are not linearized as they are when propagated with sacrificial charge. This shows that the sacrificial charge plays an important role in the linearization of the survivor radial phase space.

Table \ref{tab:Settings} summarizes the results of these optimizations. The table shows the simulation parameter ranges which yield emittances within $20\%$ of the minimum emittance.

\section{\label{Cross Section Comparison} MTE and Photoinjector Cross Section Design Effects}

In this section we briefly discuss the effect of different initial MTE values as well as the effects of the different designs for the photoinjector cross section.

\begin{figure}[htbp]
    \centering
    \includegraphics[width=220pt]{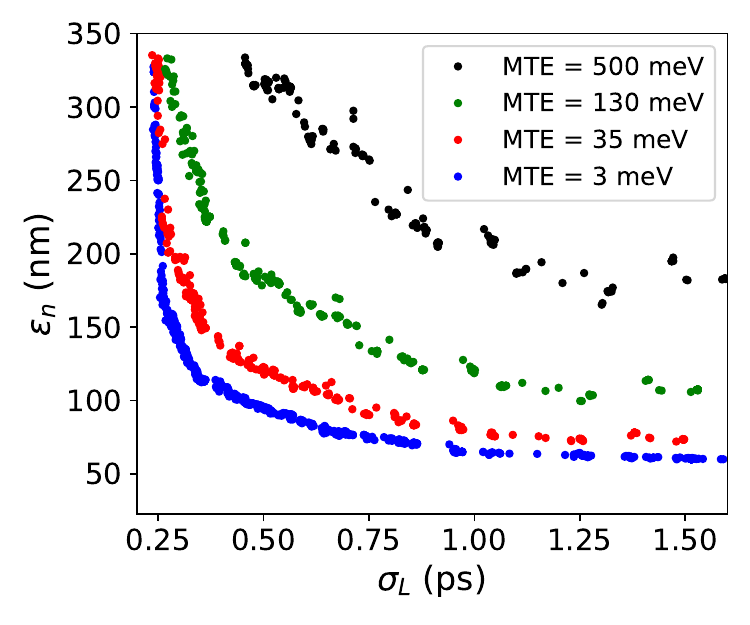}
    \caption{\label{fig:MTEfronts} Pareto fronts evaluated using different initial MTEs and using sacrificial charge and a single solenoid.}  
\end{figure}

\begin{figure}
    \centering
    \includegraphics[width=0.5\textwidth]{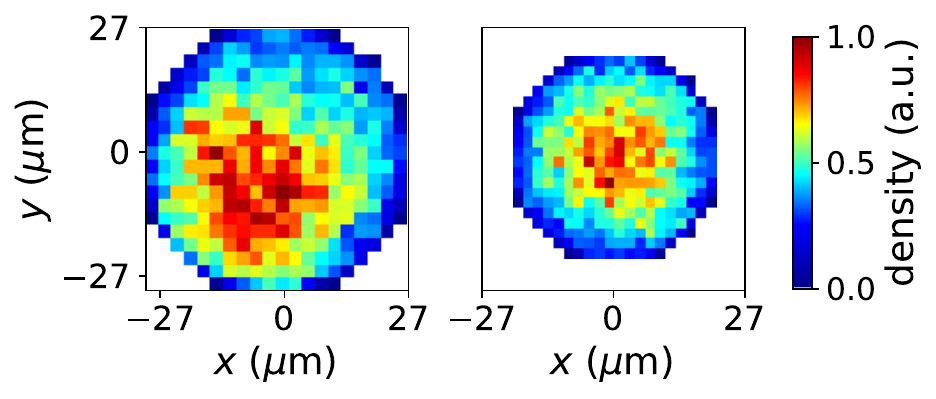}
    \caption{\label{fig:SymAsymDists} Example final survivor transverse beam distribution from optimizations with gun design with asymmetry (left) and with symmetrized design (right).}
\end{figure}

\begin{figure}[htbp]
    \centering
    \begin{subfigure}{0.5\textwidth}
        \centering
        \includegraphics[scale = 0.55]{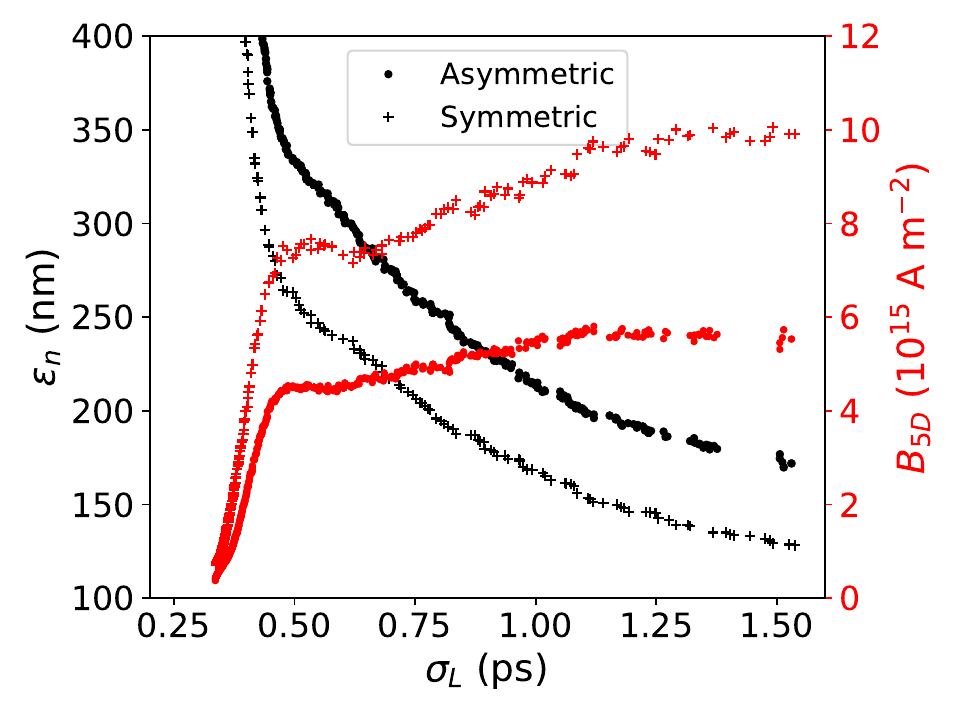}
        \caption{\label{fig:SymAsymFrontsNSC}}
    \end{subfigure}
    \begin{subfigure}{0.5\textwidth}
        \centering
        \includegraphics[scale = 0.55]{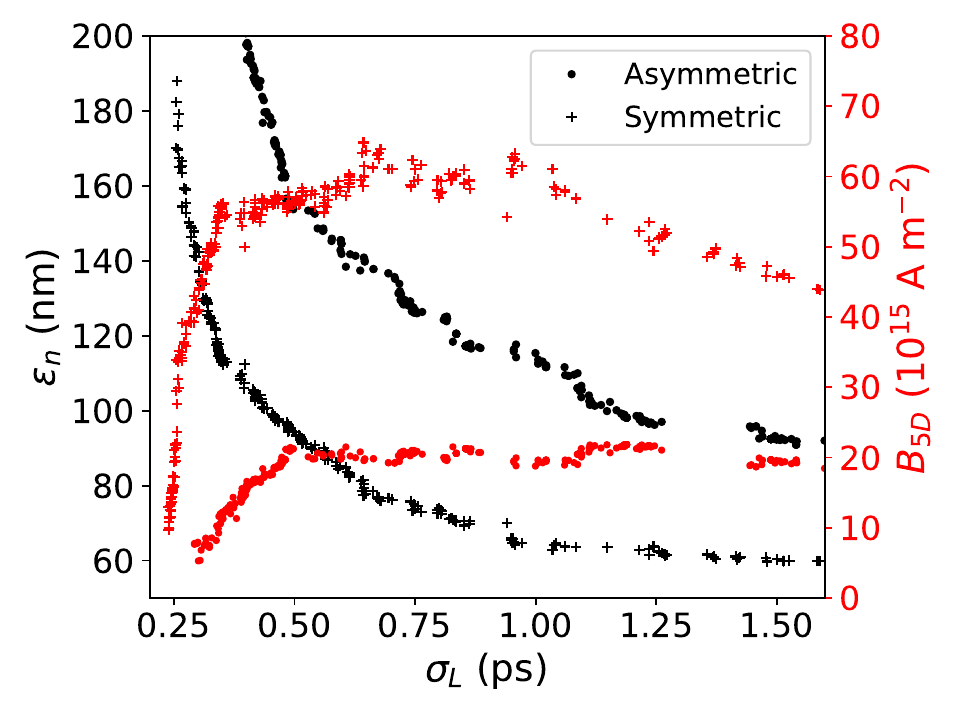}
        \caption{\label{fig:SymAsymFrontsSC}}
    \end{subfigure}
\caption{\label{fig:SymAsym} Pareto fronts from optimizations using symmetrized (crosses) and unsymmetrized (dots) gun designs. (a) Results obtained without using sacrificial charge. (b) Results using sacrificial charge with a single solenoid. The brightness is computed from the emittance and bunch length according to equation \ref{eq:Brightness}.} 
\end{figure}

Figure \ref{fig:MTEfronts} shows the fronts for different initial MTEs. These fronts were obtained by reevaluating the Pareto front from case 2, but giving the initial electron bunch different MTEs. We consider MTEs which are typical for NaKSb and metal photocathodes at room temperature. An MTE of $35~\mathrm{meV}$ corresponds to threshold photoemission from NaKSb \cite{Maxson2015-TradeoffEmittanceQE}.The MTE from beams photoemitted from NaKSb using $515~\mathrm{nm}$ light from NaKSb has been measured to be $\approx 130~\mathrm{meV}$ \cite{Maxson2015-TradeoffEmittanceQE}. A $500~\mathrm{meV}$ MTE is typical for beams from Cu photocathodes. As expected, the final emittances increase with initial MTE. However, even at the highest MTE of $500~\mathrm{meV}$ we find emittances as low as $180~\mathrm{nm}$. Importantly, these are not results from optimizations performed with different MTE values. Instead, these are the results obtained by re-evaluating with different MTE values the results previously optimized with negligible MTE.

As discussed in section \ref{Beamline Design}, the initial photoinjector design only had symmetrizing structures along one direction and an elongation in the other direction. We found this design produced a distortion in the beam phase space which led to a visible asymmetry in the transverse distribution of the beam at the location where the emittance is minimized. This can be seen in the example shown on the left in Fig. \ref{fig:SymAsymDists}. The plot on the right in Fig. \ref{fig:SymAsymDists} shows an example of the transverse beam distribution obtained with the new symmetrized photoinjector cross section showing the distortion of the beam distribution has been eliminated. Figure \ref{fig:SymAsym} compares the Pareto fronts obtained using the asymmetric photoinjector design and using the symmetrized photoinjector design. For the case without sacrificial charge, shown in Fig. \ref{fig:SymAsymFrontsNSC}, we find an up to $\approx25\%$ decrease in the emittance, corresponding to an improvement in 5D brightness of a factor of $\approx 1.7$. For the case with sacrificial charge, shown in \ref{fig:SymAsymFrontsNSC}, we find an even more significant improvement. In this case, the emittance decreases by up to $45\%$ resulting in a more than a factor of 3 improvement in the 5D brightness.

\section{Conclusion}
We performed multi-objective optimizations with the goal of minimizing the emittance of a $250~\mathrm{pC}$ electron bunch. The electron bunch is is accelerated to $\approx 6~\mathrm{MeV}$ by a 1.6 cell distributed coupling C-band photoinjector with a $240~\mathrm{MV/m}$ peak electric field on the cathode. Assuming negligible intrinsic emittance, we find the high photoinjector gradients enable emittances as low as $127~\mathrm{nm}$, surpassing the state-of-the-art for similar bunch charge and energy.

Additionaly, we performed optimizations using sacrificial charge to linearize the phase space of the core of the electron bunch with the desired $250~\mathrm{pC}$ charge. We founf the use of sacrificial charge results in the reversal of nonlinear space charge effects, further reducing the emittance down to as low as $54~\mathrm{nm}$. Furthermore, emittances as low as $100~\mathrm{nm}$ were obtained when re-evaluating the optimized settings using intrinsic emittances typical from semiconductor photocathodes.

The optimizations motivated increasing the radial symmetry of the photoinjector design. These changes resulted in substantial improvements in the achievable brightness.

Future work would study the effects of non-idealities in the initial distribution such as non-uniformities in the initial temporal profile. Further work would also investigate the effect of different accelerating gradients on the achievable brightness.

The high brightness demonstrated here due to high gradients in C-band photoinjectors and sacrificial charge promise to enable significant advances in applications such as X-ray free electron lasers and inverse Compton scattering beamlines.

\begin{acknowledgments}
This work was supported by the Los Alamos National Laboratory’s Laboratory Directed Research and Development (LDRD) Program project
20230011DR and by the Deportment of Energy,
Office of Science, Office of Basic Energy Sciences, under
award number DE-SC0020144.
\end{acknowledgments}

\bibliography{main}

\end{document}